\documentclass[12pt]{article}
\usepackage{graphics}
\usepackage{amssymb}

\def\R{{\mathbb R}}

\def\bce{\begin{center}}
\def\ece{\end{center}}

\def\frameqed{\framebox(4.25,4.25){}}
\newcommand{\bea}{\begin{eqnarray}}
\newcommand{\eea}{\end{eqnarray}}
\newcommand{\beq}{\begin{equation}}
\newcommand{\eeq}{\end{equation}}

\newtheorem{theo}{Theorem}[section]
\newtheorem{prop}{Proposition}[section]
\newtheorem{defn}{Definition}[section]

\newtheorem{cor}{Corollary} [section]
\newtheorem{rmk}{Remark}[section]
\def\bth{\begin{theo}}
\def\eth{\end{theo}}
\def\bpr{\begin{prop}}
\def\epr{\end{prop}}
\def\bdf{\begin{defn}}
\def\edf{\end{defn}}
\def\brmk{\begin{rmk}}
\def\ermk{\end{rmk}}
\def\barr{\begin{array}}
\def\earr{\end{array}}

\begin{document}

\title {Entry and exit sets in the dynamics of area preserving  H\'{e}non map
}
\author{Emilia Petrisor
\\ {\small
"Politehnica" University of Timisoara, Department of
Mathematics,}\\ {\small P-ta Regina Maria nr. 1, 1900 Timisoara,
Romania}\\{\small e-mail: epetrisor@math.uvt.ro}}
\date{}

\maketitle
\begin{abstract} In this paper we study dynamical properties of the
area preserving H\'{e}non map, as a discrete version of open
Hamiltonian systems, that can exhibit chaotic scattering.
Exploiting its geometric properties we locate the  exit and entry
sets, i.e. regions through which any forward, respectively
backward, unbounded orbit escapes to infinity.
In order to get the boundaries of these sets we prove that the right
 branch of the
unstable manifold of the hyperbolic fixed point is the graph of a
function, which is the uniform limit of a sequence of functions
whose graphs are arcs of the symmetry lines of the H\'{e}non map,
as a reversible map.

\end{abstract}
\section{Introduction}
The area preserving maps of the plane are discrete versions of
open Hamiltonian systems that can exhibit chaotic scattering
\cite{gaspard}.
Their phase space is an open and unbounded set, that is
the basic assumption of the Poincar\'{e}--recurrence theorem
\cite{arnold},
concerning their dynamics, is violated.  Hence it is natural to
wonder which points of the plane are non--wandering, and  through
which regions  do some orbits escape to infinity.\par

The simplest area preserving map of the plane
is the
quadratic H\'{e}non map \cite{henon}:
\beq
(x,y)\mapsto (y+1-ax^2,x),\eeq
where $a$ is a parameter.
 Despite its simplicity,
the area preserving H\'{e}non map  exhibits a very complex
dynamics,  described in a long list of papers appeared
during the thirty years from its definition.
Even last five years  important new results  concerning
the dynamic behaviour of this map were reported. In \cite{meiss},
for example,
 Meiss presents a theoretical support for the study of transport
 through the resonance zone associated to the hyperbolic fixed point
 of the H\'{e}non map,
 as well as, for a numerical tool
designed to compute the measure of the bounded orbits of the H\'{e}non
map. Homoclinic bifurcations for this map are discussed in
\cite{sterling}, and the computation of its periodic orbits using
anti--integrable limit in \cite{sterling0}.  A method for computation
of the self--rotation number for its orbits is given in
\cite{dullin3}.\par

In this paper we study the  H\'{e}non map from the point of view of open
conservative systems, i.e. conservative systems whose phase space is unbounded.
We  prove the existence of an unbounded forward
(backward)  invariant set, called {\it exit set} ({\it entry set})
through which any forward (backward) unbounded orbit escapes to  infinity.
These sets are very useful in the study of transport properties
of H\'{e}non maps \cite{meisst, easton}, as well as of chaotic scattering in their
dynamics \cite{gaspard,lau}.\par
In order to locate these regions for the H\'{e}non map, we study its
geometric properties, and prove that the right branch of the
unstable manifold of the hyperbolic fixed point is the graph of a
function, which is the uniform limit of a sequence of functions
whose graphs are arcs of the symmetry lines of the H\'{e}non map,
as a reversible map.

\section{Regions of different dynamical behaviour for H\'{e}non maps}
Some dynamical properties of polynomial mappings with polynomial inverse
are studied in \cite{fmilnor}. These maps are called {\it Cremona maps}
and they form a group $G$.
 Planar diffeomorphisms $g_\delta\in G$ of the form
\beq\label{henong}
(x,y)\stackrel{g_\delta}{\longmapsto}(y, p(y)-\delta x),\eeq
where $p$ is a polynomial function of degree at least two, and
$\delta\neq 0$ is the constant Jacobian determinant of the map $g_\delta$,
 are
called in \cite {fmilnor} generalized H\'{e}non transformations.\par

For $\delta=1$ in (\ref{henong})  we
get polynomial area preserving maps with polynomial inverse.\par

By \cite{fmilnor} any Cremona map $F$ of prime degree $d$ is affinely
conjugate either to a generalized H\'{e}non map or to an elementary
transformation $e(x,y)=(ax+p(y),by+c)$, $a\,b\neq 0$. In the first
case the generalized H\'{e}non map can have the polynomial $p$
in the normal form $p(x)=\pm x^d+\mbox{terms of degree} \leq
d-2$.\par

For our purposes it is more appropriate to study the class of
area preserving generalized H\'{e}non maps of the form
$g_1^{-1}$, i.e. maps defined by:
\beq\label{myhenon}
(x,y)\stackrel{H_d}{\mapsto}(-y+p(x,\mu),x),\eeq
where $d=2,3$ is the degree of the polynomial family $p(x,\mu)$.
Because in this case $d$ is a prime number, a particular choice of
the polynomial family $p(x,\mu)$ does not affect the generality
of results. Namely, we study   the quadratic H\'{e}non map
$H_2(x,y)=(-y+\mu x +x^2,x)$, and in a forthcoming paper
 the cubic H\'{e}non maps
$H_3^{\pm}(x,y)=(-y\pm x^3+\mu x+\nu,x)$, as open systems.\par

In our study we exploit the reversibility property of the H\'{e}non
map.
The area preserving maps (\ref{myhenon}) are reversible with respect to
the involution $R:\R^2\to\R^2$, $R(x,y)=(y,x)$, i.e.
$H_d^{-1}=R\circ H_d \circ R$. Hence the map $H_d$ factorizes as
$H_d=I\circ R$, where $I(x,y)=(-x+p(y,\mu), y)$ is also an involution.\par


Recall some properties of the reversible systems we use in our
approach
 (a survey on the dynamics of reversible
systems is presented in \cite{lamb}).\par
 The relation $H_d^{-1}=R\circ H_d\circ R$ ensures that
$H_d^{-n}=R\circ H_d^{n}\circ R$ also holds, for any $n\in\mathbb{Z}$.
Hence $H_d^n$ is  $R$-reversible, too. Denote by $I_n$ the involution
defined by $I_n=H_d^n\circ R$, $n\in \mathbb{Z}$,
 and by $\Gamma_n$ its fixed point sets, called $n$--symmetry
 line:
 \beq\label{defgammak}
 \Gamma_n=\{(x,y)\,\vert\,(H_d^n\circ R)(x,y)=(x,y)\}
 \eeq
  If
$R$ is a $C^1$--orientation reversing involution, then $\Gamma_0$ is
nonempty, and moreover it is a $C^1$-curve in $\mathbb{R}^2$ having no
self--intersections.\par

The symmetry lines $\Gamma_k$ are transformed by $H_d^n$  into other
 symmetry lines in the following way:
\beq\label{trsymL} H_d^n(\Gamma_k)=\Gamma_{2n+k},\,\,\forall\,\, n,k\in\mathbb{Z}\eeq
For the H\'{e}non map (\ref{myhenon}),
$R$ and $I$ are orientation reversing involutions and
their fixed point sets are, respectively $\Gamma_0: x=y$,
$\Gamma_1: x=p(y,\mu)/2$. An $R$--invariant periodic orbit of
the map $H_d$
is called symmetric orbit. The symmetric fixed points of the map $H_d$
lie at the intersection of the two basic symmetry lines $\Gamma_0$
and $\Gamma_1$.\par
H\'{e}non maps being discrete versions of open Hamiltonian
systems, it is very important to
locate the non--wandering set for such a map, and the region
through which forward unbounded orbits escape to infinity.
  A point $z\in\mathbb{R}^2$
is non--wandering for the H\'{e}non map $H_d$, if for any
neighbourhood $V$ of $z$, there exists an integer  $n>0$,
such that $H_d(V)\cap V\neq\emptyset$. In other words,  after some time
$n>0$,
the orbit of  a non--wandering point $z\in V$, comes back into the neighbourhood
$V$.
\par

 The set of non-wandering points of the H\'{e}non maps $H_d$,
 denoted $\Omega(H_d)$,
is proved to be contained in a particular region of the phase space
\cite{fmilnor}.
In this paper we locate the entry and the exit set for the quadratic
 H\'{e}non map.\par
If a generalized H\'{e}non map has no fixed points, then every orbit
is unbounded \cite{dullin1}. In the case when (\ref{myhenon}) has fixed
points, then there is a box $B$, such that any orbit outside it is
either forward unbounded or backward unbounded or both. This box is:
\begin{equation}
B=\{(x,y)\,\vert\,\, \vert x\vert <M, \vert y\vert <M\},\end{equation}
where $M$ is the largest of the absolute values of the roots of the
equation $\vert p(x,\mu)\vert-2\vert x\vert=0$. Hence bounded orbits
lie inside the box.\par
The quadratic family $H_2(x,y)=(-y+\mu x+x^2,x)$ has two $R$--symmetric fixed
points: $z_e=(0,0)$ and  $z_h=(2-\mu, 2-\mu)$. Denote $x_h:=2-\mu$.\par
We study the  maps corresponding to the parameter
 $\mu\in(-2,2)$. In this range
the origin is elliptic, while the second fixed point is hyperbolic.
 The value $M$ in the definition of the box $B$ is, in this case
$M=x_h$. The bounded orbits of the
quadratic H\'{e}non map, corresponding to $\mu\in(-2,2)$,  are periodic orbits,  all orbits
filling densely invariant circles surrounding the elliptic fixed point
whose multipliers are $\lambda,\overline{\lambda}=e^{\pm 2\pi i
\omega}$,
with $\omega\neq 1/3, 1/4$, as well as orbits on the cantori
(former invariant circles, which have broken down).\par
Beside the orbits  outside the box $B$, it is possible
to exist unbounded orbits which enter it, spend some time in  $B$,
and then are scattered to infinity.
Chaotic scattering in the dynamics of area preserving
maps of the plane was revealed for the quadratic map $f(x,y)=(\overline{x},\overline{y})$ \cite{lau}:
 \beq\label{scattermap}\barr{lll}
\overline{x}&=&[x-(x+y)^2/4]a\\
\overline{y}&=&[y+(x+y)^2/4]a^{-1}\earr
 \eeq
 \noindent where $a>1$ is a parameter.

Because any quadratic area preserving map of the plane, with
nontrivial dynamics, is affinely conjugated to the H\'{e}non map  $H_2$,
it results from the study of the  map (\ref{scattermap}), that
 in some range of the parameter $\mu$ of the H\'{e}non map
  can also exist a subset $A$ in the box $B$,
such that almost every point in $A$ belongs to a trajectory that
eventually leaves $B$ (it is scattered to infinity). The stable set
$S(A)$ (unstable set $U(A)$) is the set of points in $A$ whose forward
(backward) orbit stays in A. The set $C=S(A)\cap U(A)$ is
$H_2$--invariant. If this set is a Cantor set which contains a
 dense orbit in $C$,  of positive Lyapunov exponent,  then
$C$ is called {\it chaotic saddle}. Hence the orbits of points on a
chaotic saddle are also bounded.
 $A\setminus C$ is called {\it scattering region}. A trajectory
 entering the scattering region typically spends a finite amount of
 time near the chaotic saddle, and then exit it, as well as the box $B$,
  being scattered to infinity.
   The scattering process is called chaotic if the motion near the
 chaotic saddle is chaotic.
  Chaotic scattering is a manifestation
 of transient chaos in open Hamiltonian systems. In order to
 quantify the scattering process   in the dynamics of
 area preserving maps one defines  the {\it exit--time
 function},
 and the {\it time--delay function} \cite{gaspard}. The exit--time function
  associates to an initial
 position in $A$ the minimal time needed to reach the exit set,
 while
 the time--delay function associates to a point in the entry set
 the minimal time to reach the exit set.
Each of these functions gives information on the time
  that the moving  particle spends near the chaotic saddle,
 bouncing between its points. Because the points of the chaotic
 saddle do not leave $A$, the exit--time function has
a Cantor set of singularities.\par
As far as we know, the entry and  exit sets
of the polynomial  area preserving
 maps, analyzed from the point of view of chaotic scattering, have not been
 located,
 and as a consequence
  the exit--time of a scattered particle $z_0$ was associated in a
subjective way, namely as the the minimal time $n$ such that the $n^{th}$ iterate
of the map applied to $z_0$ is outside a large disk centered at the origin \cite{lau, gaspard}.
Moreover the time--delay function could not be evaluated because
the entry set was unknown.\par
Next we show that for the quadratic H\'{e}non map we can determine
analytically an exit set (entry set) of the system, i.e. an unbounded forward
 (backward) invariant
set $E$ ($F$), such that any forward (backward) unbounded orbit enter $E$
($F$),
and as a consequence goes to infinity through this set.\par

The entry and exit sets are also useful in the study of transport
through the resonance region, bounded by segments of stable and
unstable manifolds of the hyperbolic fixed point of the H\'{e}non
map \cite{meisst, meiss, easton}.
\section{The exit and entry set of the quadratic H\'{e}non map}
In order to detect the exit and entry sets of the H\'{e}non map
we study its geometric properties.
 For simplicity we
drop the index $2$ from its name $H_2$.
\bpr\label{defsetE}  The H\'{e}non map $H(x,y)=(-y+\mu x+x^2,x)$,  $\mu\in(-2,2)$, has an
unbounded forward invariant set
$E_0\subset\{(x,y)\in\mathbb{R}^2\,\vert\,\, x>0,y>0\}$, i.e. $H(E_0)\subset E_0$.
Its boundary $\partial E_0$ is ${\cal L}\cup \Gamma_0^h$, where ${\cal L}$ is
the curve parameterized by $(X(a), Y(a))$,
$a\in (0,1]$, with
\begin{equation}\begin{array}{l}
X(a)= \displaystyle\frac{a^2-a\mu+1}{a}\\
Y(a)=aX(a),\end{array}\end{equation}
and $\Gamma_0^h=\{(x,y)\,\vert \, y=x,\, x\geq x_h\}$.
\epr
Proof\par

\begin{figure}[p]
\begin{center}
\includegraphics{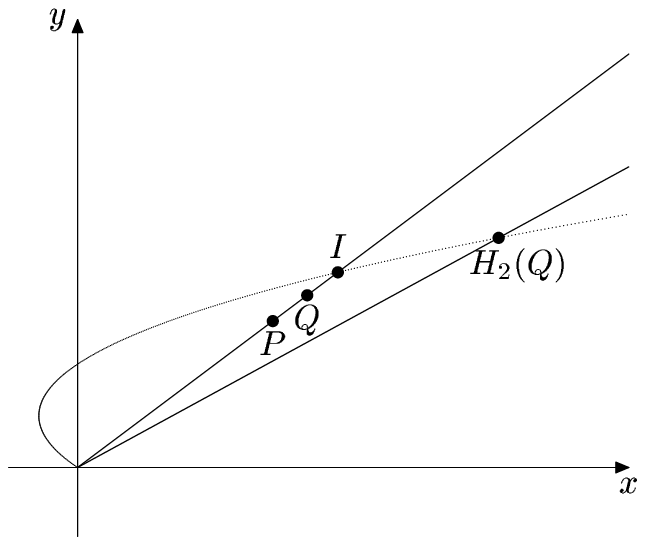}
\end{center}
\caption{\label{exitdem} The action of the map $H$ on a semi--line $L_1(a)$, $a\leq 1$.}
\end{figure}

Let $L(a)$ be the infinite radius (semi--line) $y=a\, x$,
  starting from the elliptic
fixed point $(0,0)$.
 Its image under
$H$ is an arc of parabola, $x=y^2+(\mu-a)y$. Only the radii
$L_1(a)$
defined by $x>0$, $a>0$, intersect their images $H(L_1(a))$ at the
points
$
I(a)=\left(\displaystyle\frac{a^2-a\mu+1}{a^2},
\frac{a^2-a\mu+1}{a}\right)$ (Fig.\ref{exitdem}).\par

The point $I(a)$ is the $H$--image of the point
$P(a)=\left(\displaystyle\frac{a^2-a\mu+1}{a},a^2-a\mu+1\right)$.
For any $\mu\in(-2,2)$, $I(a)$, with $a>0$, belongs to the set
 $S=\{(x,y)\,\vert \,x>0,y>0\}$, as well as the sub--arc of
  the parabola $x=y^2+(\mu-a)y$
corresponding to $y>y_{I(a)} $. This suggests to look for a forward
invariant set  in $S$.\par

Any point $Q(x_0,a\, x_0)\in L_1(a)$, $0<a\leq 1$, with $x_0> x_{P(a)}$ is
mapped by $H$ to the point $(x_1, a'\,x_1)\in L_1(a')$
 (see Fig.\ref{exitdem}), with
\begin{equation}\label{aprim}
 a'(x_0,a)=\displaystyle\frac{1}{x_0+\mu-a}<a
\end{equation}
Let us show that the set points
\begin{equation}
E_0= \{(x,a\,x)\,\vert\, x>\displaystyle\frac{a^2-a\mu+1}{a}, 0<a\leq
1\}\end{equation}
is forward invariant.\par
For, take a point   $(x_0,a\,x_0)$  in $E_0$.
By (\ref{aprim}),
$\displaystyle\frac{a'^2-a'\mu+1}{a'}=\frac{1}{x_0+\mu-a}+x_0-a$.
 From $a'<a$, we have
$\displaystyle\frac{a'^2-a'\mu+1}{a'}<x_0$. For $0<a\leq 1$,
$\displaystyle\frac{a^2-a\mu+1}{a}>1-\mu+a$. Hence
$x_0>1-\mu+a$, i.e. $x_0+\mu-a>1$, and thus $x_1>x_0$ or
equivalently $x_1>\displaystyle\frac{a'^2-a'\mu+1}{a'}$. This means
that
$(x_1,a'\,x_1)=H(x_0,a\,x_0)\in E_0$, for any $(x_0,a\,x_0)\in
E_0$. \frameqed\par\vskip 0.5cm
\brmk\label{evrot} In canonical polar coordinates, a point $(r_0,\theta_0)\in E_0$
is mapped by $H$ to a point $(r_1,\theta_1)$, with
$\theta_1<\theta_0$. Any point $(r_0,\theta_0)\in (0,\infty)\times [\pi/2,2\pi]$,
 is mapped to a point
$(r_1,\theta_1)$, where $\theta_1$ is obtained from $\theta_0$
by a positive translation (rotation) on the circle.\ermk

\bpr\label{propfo} The forward orbit $(x_n,y_n)=H^n(x_0,y_0)$, $n\in\mathbb{N}$,
of a point $(x_0, y_0)\in E_0$ has the property that both sequences
 $(x_n), (y_n)$ are increasing and unbounded.\epr
Proof\par
Projection onto the first factor of the set $E_0$ is
$(x_h,\infty)$. Hence $x_0>x_h$.
From the proof of the  Proposition \ref{defsetE} it results  that
the sequence $(x_n)$ is increasing. Moreover
\beq\label{secord}
x_{n+1}+x_{n-1}=p(x_n),\eeq
 with $p(x)=\mu x+x^2$.
The equation (\ref{secord}) is the H\'{e}non map expressed as the second
order difference equation. The sequence $(x_n)$ is unbounded,
because otherwise it converges to a point $x^*$
satisfying, by (\ref{secord}),   the fixed point
equation associated to the H\'{e}non map, $p(x^*)=2x^*$, i.e.
$x^*\notin
\mbox{pr}_1(E_0)$.\par
 In a similar way one proves that the sequence $(y_n)$ is
increasing and unbounded ($y_n$ satisfies the same second order
difference equation). \frameqed\par\vskip 0.5cm
 As a consequence,
in canonical polar
coordinates, the orbit $(r_n,\theta_n)$, $n\in \mathbb{N}$, of a  point
 $(r_0,\theta_0)\in E_0$,
has $r_n=(x_n^2+y_n^2)/2\to\infty$, i.e. the orbit goes to
infinity.\par

Let $\lambda^u>1$ be the expanding multiplier of the hyperbolic fixed
point $z_h$, and
 $v=(\lambda^u,1)$ a corresponding eigenvector. The branch of the
 unstable manifold $W^u(z_h)$ having the direction and the sense of $v$
 for the
tangent semi-line $l$ at $z_h$
  is denoted by $W^u_+(z_h)$, and called the right branch of the
 unstable manifold.

\bpr The right branch $W^u_+(z_h)$ of the unstable manifold
of the hyperbolic fixed point $z_h$
is included in the set $E_0$. \epr
Proof\par

A simple computation shows that the  segment $l=\{z\in\mathbb{R}^2\vert
z=z_h+tv, 0<t\leq 1\}$ of the tangent at $z_h$, to the unstable
manifold,
is included in the  set $E_0$.
Hence, there is a $\delta>0$ such that an arc
of the local unstable manifold
$W^u_{loc}(z_h)=\{z\in\mathbb{\R}^2 \vert\, H^{-n}(z)\in B_{\delta}(z_h),
  \forall n\geq 0\}$ is included in $E_0$ ($B_\delta(z_h)$ is the $\delta$--ball
   centered at $z_h$).
    Denote by $L^u$ this arc. The right branch
   of unstable manifold is then
 $W^{u}_+(z_h)=\bigcup\limits_{n\geq
 0}^{}H^n(L^u)$, and by the forward invariance of $E_0$, it is
  included in this set. \frameqed\par\vskip 0.5cm

The H\'{e}non map $H$ being reversible, its inverse is conjugated
with $H$ by the involution $R(x,y)=(y,x)$. Hence the set $F_0=R(E_0)$ is a
backward invariant subset, i.e. $H^{-1}(F_0)\subset F_0$. Its boundary is
$\partial F_0={\cal L}'\cup
\Gamma_0^h$, where ${\cal L}'=R({\cal L})$.
Furthermore, the right branch, $W^s_+(z_h)=R(W^u_+(z_h))$, of the  stable manifold $W^s(z_h)$,
 is included in the set $F_0$.\par

 Next we  determine the maximal forward, respectively
 backward, invariant set in $S=\{(x,y)\,\vert\, x>0,y>0\}$. For,
 we exploit the fact that $\Gamma_0^h=E_0\cap F_0$, i.e. orbits
starting from  points
$z_0=(x_0,x_0)$, with $x_0>x_h$, are unbounded in both directions.
Moreover, $\Gamma_{-2n}^h:=H^{-n}(\Gamma_0^h)\subset F_0$,
and   $\Gamma_{2n}^h:=H^{n}(\Gamma_0^h)\subset E_0$, $\forall\,\, n\in\mathbb{N}$.

 Because $\Gamma_0^h\subset E_0$ it follows that the sequence of
 sets $(E_n)$, $n\in\mathbb{N}$, with $\partial E_n={\cal L}\cup \Gamma_{-2n}^h$,
   is a sequence of forward
 $H$--invariant sets. We prove that it is an increasing sequence
 and its limit has as a partial boundary the right branch,
 $W^s_+(z_h)$, of the stable manifold of the hyperbolic fixed point
 $z_h$.\par
\bpr The right branch, $W^s_+(z_h)$, of the stable manifold  is the graph
of an increasing  $C^{1}$--function
$\varphi:[x_h,\infty)\to[x_h,\infty)$, which is the uniform limit
of a sequence of increasing
$C^{1}$--functions $\varphi_n:[x_h,\infty)\to[x_h,\infty)$,
$n\in\mathbb{N}$, and:\\
i) $\varphi_n(x_h)=x_h$, $\forall\,\, n\in \mathbb{N}$;\\
ii) $\varphi_n(x)>\varphi_m(x)$, $\forall\,\, n>m$, and $x>x_h$;\\
iii)  the graph of the function $\varphi_n$
is the partial boundary $\Gamma_{-2n}^h$ of the forward invariant
set $E_n$.\epr

Proof\par

Adapting a result from \cite{caomao} to the case of
 area preserving H\'{e}non map,
 we have  that
the right branch $W^u_+(z_h)$ of the unstable manifold
is the graph of an increasing
$C^1$--function $g:[x_h,\infty)\to(x_h,\infty)$. The right branch of
stable manifold $W^s_+(z_h)=R(W^u_+(z_h)$ being the symmetric of
 $W^u_+(z_h)$
with respect to the line $y=x$,  is the graph of the function
$\varphi=g^{-1}$, and this function is also increasing.\par

The partial boundary $\Gamma_0^h$ of the set $E_0$ is the graph of
the function $\varphi_0=\varphi_0^{-1}$, $\varphi_0(x)=x$, $x\geq x_h$. Its
image under $H$ is the set of points
$H(x,\varphi_0^{-1}(x))=(-\varphi_0^{-1}(x)+p(x),x)$, with $x\geq x_h$,
and
$p(x)=\mu x+x^2$. Denote by $\varphi_1(x):=p(x)-\varphi_0^{-1}(x)$.
Obviously, $\varphi_1(x_h)=x_h$, and  $\varphi_0(x)<\varphi_1(x)$, $\forall\,\,
x>x_h$.
The derivative of $\varphi_1$ has the property:
\beq\label{derm1}
\varphi'_1(x)>1, \,\,\forall\,\, x\geq x_h\eeq
Hence it is an increasing, that is invertible function, and the graph
 of its inverse is
$\Gamma_2^h$,
while the graph of $\varphi_1$ is $\Gamma_{-2}^h=R(\Gamma_2^h)$.\par
$\Gamma_4^h$ is thus the $H$--image of the graph of the function  $\varphi_1^{-1}$,
i.e. the set of points $H(x,\varphi_1^{-1}(x))$
$=(p(x)-\varphi_1^{-1}(x),x)$,
$x\geq x_h$. Denote by $\varphi_2$ the function defined by
$\varphi_2(x):=p(x)-\varphi_1^{-1}(x)$. $\varphi_2(x_h)=x_h$.
Because $\varphi_0(x)<\varphi_1(x)$, $\forall\,\,
x>x_h$, is equivalent to $\varphi_1^{-1}(x)<\varphi_0^{-1}(x)$, $\forall\,\,
x>x_h$, we get from the definition of the functions $\varphi_i$, $i=1,2$,
that $\varphi_2(x)-\varphi_1(x)=\varphi_0^{-1}-\varphi_1^{-1}>0$.
 The property
(\ref{derm1}) and the fact $p'(x)>2$, for any $x\geq x_h$,
 ensures that $\varphi'_2(x)>1$, $\forall
x\geq x_h$. Hence the graph of $\varphi_2^{-1}$ is $\Gamma_4^h$, and
the graph of $\varphi_2$ is $\Gamma_{-4}^h$. By induction, we
get that the partial boundary $\Gamma_{-2n}^h$ of
the set $E_n$, is the graph of an increasing $C^{1}$-function
$\varphi_n:[x_h,\infty)\to[x_h,\infty)$, defined by:
\beq\label{recfun}
\varphi_n(x):=p(x)-\varphi_{n-1}^{-1},\,\,\forall\,\, x\geq x_h,\eeq
and it fulfills the conditions: $\varphi_n(x_h)=x_h$,
$\varphi_{n}(x)>\varphi_{n-1}(x)$, $\forall\,\, x>x_h$,
 and $\varphi'_n(x)\geq 1$.\par

 No arc $\Gamma_{-2k}^h$  can intersect the stable manifold $W^s(z_h)$ in $F_0$.
To prove this property, suppose  that there exists a
point $z\in W^s_+(z_h)\cap\Gamma_{-2k}^h$, $z\neq z_h$.
 Hence $\lim_{n\to\infty}
H^{n}(z)=z_h$.  Because $\Gamma_{-2k}^h\subset\Gamma_{-2k}$, by
(\ref{defgammak})
we have $(H^{-2k}\circ R)(z)=z$. At the same time, $Rz\in W^u(z_h)$, and
$\lim_{n\to\infty}H^{-n}(Rz)=z_h$ or equivalently
$\lim_{n\to\infty}H^{-n+2k}(H^{-2k}R(z))=z_h$. Hence the orbit of $z$
is a homoclinic orbit to $z_h\notin F_0$, which contradicts the fact
that $z$ is a point in  the backward invariant set $F_0$.\par
Because $W^s_+(z_h)$ cannot intersect any arc of symmetry line
$\Gamma_{-2n}$ in $F_0$, we have that $\varphi_n(x)<\varphi(x)$, for any
$x>x_h$. Hence the sequence of functions $(\varphi_n)$ converges
uniformly to an increasing $C^1$--function
$\psi:[x_h,\infty)\to[x_h,\infty)$. By (\ref{recfun}) this function
has the property that $p(x)-\psi(x)=\psi^{-1}(x)$. Hence a point
 $(x_0,\psi(x_0))$ of
its graph is mapped by the H\'{e}non map to
$H(x_0,\psi(x_0))=(p(x_0)-\psi(x_0),x_0)=(\psi^{-1}(x_0), x_0)$. Denoting
$x_1=\psi^{-1}(x_0)$, we have $H(x_0,\psi(x_0))=(x_1,\psi(x_1))$, i.e. the
graph of the function $\psi$ is forward invariant under the action of the map $H$. If
$(x_n,\psi(x_n))=H^n(x_0,\psi(x_0))$, then
$(x_{n-1},\psi(x_{n-1}))=H^{-1}(x_n,\psi(x_n))$. By Proposition \ref{propfo}
and the reversibility of the system we get that $x_{n-1}>x_n$, $\forall\,\,
n>0$,
 and
$x_n\to x_h$. Hence $H^n(x_0,\psi(x_0))\to (x_h,x_h)$, as
$n\to\infty$, that is, the graph of $\psi$ is the right branch of
the stable manifold of the fixed point $z_h$, i.e. $\psi=\varphi$.
\frameqed
\begin{cor} The sequence of forward invariant sets $E_n$ having the
boundary
$\partial E_n={\cal L}\cup\Gamma^h_{-2n}$ is an increasing sequence and
its limit $E=\cup_{n\geq 0}E_n$ is forward invariant, too, and
has the boundary $\partial
E={\cal L}\cup W^s_+(z_s)$.\par
 The set $F=R(E)$ is backward invariant,
and $\partial F={\cal L}'\cup W^u_+(z_s)$.\par
The set of points $G=E\cap F=\{(x,y)\,\vert\, x>x_h,
\varphi^{-1}(x)<y<\varphi(x)\}$ is $H$--invariant, i.e. $H(G)=G$.
\end{cor}
Because the orbit of a point in $G$ is unbounded in both
directions, the exit and entry sets  of dynamical interest for transport
processes, as well as for chaotic scattering are
$E'=E\setminus G$, respectively $F'=F\setminus G$.
The boundary for $E'$ is $\partial E'={\cal L}\cup
W^u_+(z_h)$, while for $F'$ is $\partial F'={\cal L}'\cup
W^s_+(z_h)$.\par
\begin{figure}[p]
\begin{center}
\includegraphics{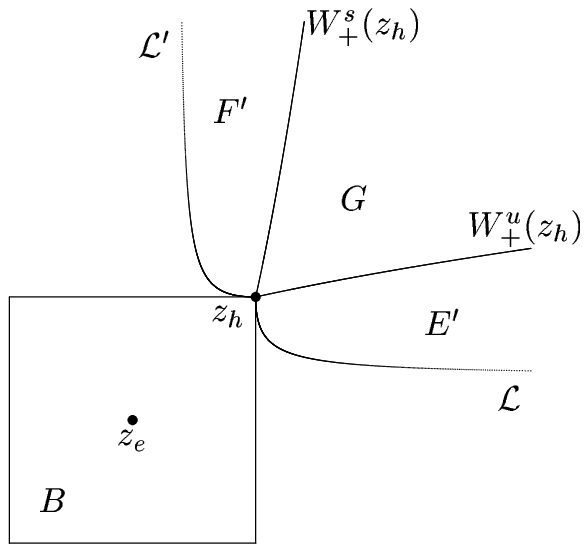} \includegraphics{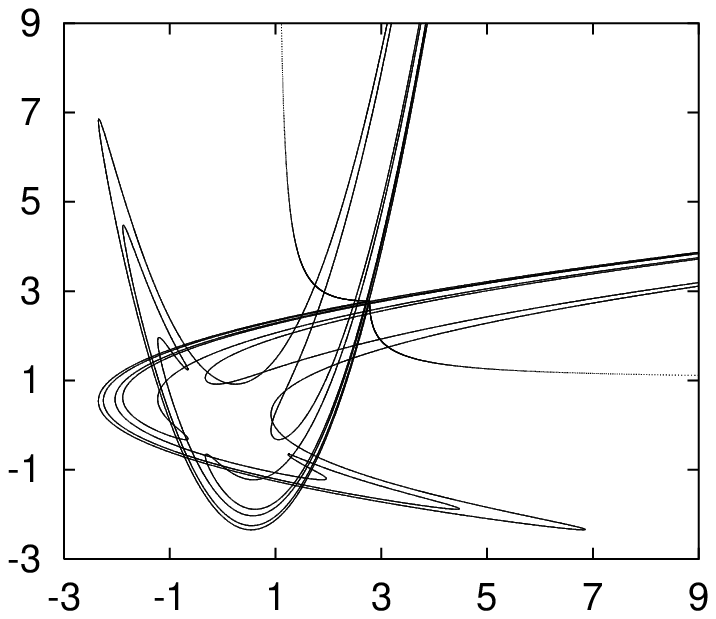}
\end{center}
\caption{\label{henonex} a) Regions of different dynamical behaviour of the H\'{e}non
map,
corresponding to $\mu=-0.78$; b) The invariant manifolds of the hyperbolic
fixed point.  Within the sets $E'$ ($F'$)
one can see foldings of the left branch of the unstable (stable) manifold of the point $z_h$.}
\end{figure}

 The position of these sets in the phase space of the map is illustrated
in Fig.\ref{henonex}a, and the position of the invariant manifolds
of the hyperbolic fixed point,
with respect to these regions is shown in Fig.\ref{henonex}b.

By straightforward computation we get that the preimage of the curve
 $\cal L$, as well as the image of ${\cal L}'$,  intersects the box
 $B$. Hence there exist in the box $B$, points $z$, such that
 $t^+(z)=\min\{n\geq 0\,\vert \, H^n(z)\in E'\}=1$, respectively
 points $z'$ with
 $t^{-}(z')=\min\{n\geq 0\,\vert \, H^{-n}(z)\in F'\}=1$.
 This means that indeed $E'$, $F'$ are exit, respectively entry,
 set for the H\'{e}non map, as a map having a scattering region.\par
 Knowing these sets, a typical scattering experiment
consists in launching particles $z\in F'$ and observing their
trajectory. If such a trajectory enters the box $B$ and then exits
and reaches
$E'$,  the time--delay is the time it takes to a particle to go
from the launching position until it reaches the exit set $E$. To a
particle $z\in F$, which do not enter $B$, one associates the time
$0$. The existence of particles in $F'$ that do not enter $B$ is
ensured by the fact that $H({\cal L}')$ intersect
the complementary set of $B$, not only the set $B$.\par
\begin{figure}[p]
\begin{center}
\includegraphics{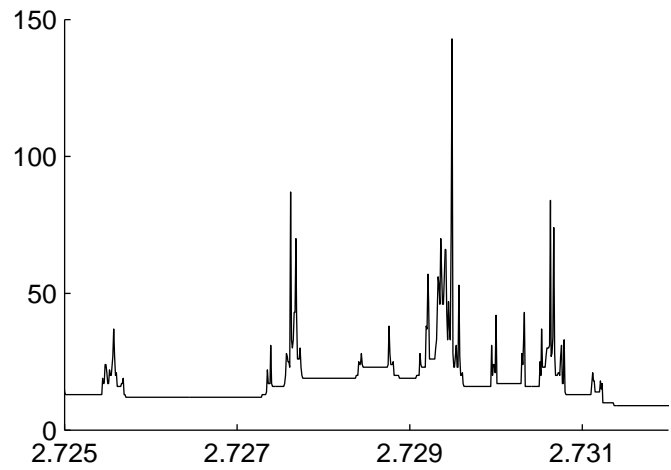}
\end{center}
\caption{\label{tmdelay1} The time--delay function defined on the segment
 $y=3.61$, $x\in [2.66, 2.74]$ included in the entry set $F'$ of the
 H\'{e}non map corresponding to $\mu=-0.75$.}
\end{figure}

In Fig.\ref{tmdelay1} is illustrated the time--delay function
of points in the entry set $F'$ belonging to the segment
 $y=3.61$, $x\in [2.66, 2.74]$.

Note that the points of the curve $\cal L$ are mapped
to points in $E'$. Namely, the points on $\cal L$ are the
points $P(a)$ in the Proposition \ref{defsetE},  and they are applied to
points $I(a)\in E'$, i.e. to points on the same semiline
of slope $a$ (see Fig.\ref{exitdem}). Moreover, a simple
analytical computation
shows that points on the semi--line $y=0$, $x> x_h$, as well as points
in the complement of the box $B$,
lying between this semi--line and the curve $\cal L$, are applied in few iterations
of the map into the set $E'$.  Taking also into account the Remark
\ref{evrot},
we conclude that any point from the exterior of the set $B$, having a forward
unbounded orbit,  reaches the
exit set $E$, in a small number of iterations.
\section{Conclusions} Studying geometric properties of the
quadratic area preserving H\'{e}non map, we have located in the
phase space its exit and entry sets. The knowledge of these sets
allows a more rigorous study of transport properties, as well as
of chaotic scattering processes in the dynamics of this map.
It appears that unlike classical Hamiltonian systems, exhibiting chaotic
scattering, in the dynamics of the H\'{e}non map, the particles
cannot enter  the scatterer  from any direction of the phase
space, and cannot exit from the system in any direction. The entry (exit) set is located
 in a neighbourhood of a branch of
stable (unstable) manifold of the hyperbolic fixed point of the
map.\par
In a forthcoming paper we determine these sets for generalized
H\'{e}non maps of degree three, which have either an elliptic fixed point
and two hyperbolic fixed  points, or a single elliptic point, but a two
periodic hyperbolic orbit.
\section{Acknowledgements} This work is supported by
NATO Linkage Grant PST.CLG.977397, and it was completed during our visit
 to Euratom--CEA Cadarache, France.  We  thank J. Misguich
 for his support and useful discussions.

\pagebreak

\newpage

\end{document}